\documentclass[8.5pt,twoside,twocolumn]{article}
\usepackage[super,sort&compress,comma]{natbib}
\usepackage{mhchem}
\usepackage{times,mathptmx}
\usepackage{sectsty}
\usepackage{balance}
\usepackage{graphicx} 
\usepackage{lastpage}
\usepackage[format=plain,justification=raggedright,singlelinecheck=false,font=small,labelfont=bf,labelsep=space]{caption}
\usepackage{fancyhdr}
\begin{document}

\thispagestyle{plain}
\fancypagestyle{plain}{
\renewcommand{\headrulewidth}{1pt}}
\renewcommand{\thefootnote}{\fnsymbol{footnote}}
\renewcommand\footnoterule{\vspace*{1pt}%
\hrule width 3.4in height 0.4pt \vspace*{5pt}}
\setcounter{secnumdepth}{5}

\makeatletter
\def\subsubsection{\@startsection{subsubsection}{3}{10pt}{-1.25ex plus -1ex minus -.1ex}{0ex plus 0ex}{\normalsize\bf}}
\def\paragraph{\@startsection{paragraph}{4}{10pt}{-1.25ex plus -1ex minus -.1ex}{0ex plus 0ex}{\normalsize\textit}}
\renewcommand\@biblabel[1]{#1}
\renewcommand\@makefntext[1]%
{\noindent\makebox[0pt][r]{\@thefnmark\,}#1}
\makeatother
\renewcommand{\figurename}{\small{Fig.}~}
\sectionfont{\large}
\subsectionfont{\normalsize}

\fancyfoot{}
\fancyfoot[RO]{\footnotesize{\sffamily{ \hspace{2pt}\thepage}}}
\fancyfoot[LE]{\footnotesize{\sffamily{\thepage }}}
\fancyhead{}
\renewcommand{\headrulewidth}{1pt}
\renewcommand{\footrulewidth}{1pt}
\setlength{\arrayrulewidth}{1pt}
\setlength{\columnsep}{6.5mm}
\setlength\bibsep{1pt}
\twocolumn[
  \begin{@twocolumnfalse}
\noindent\LARGE{\textbf{Tunable interactions between paramagnetic colloidal particles driven in a modulated ratchet potential}$^\ddag$}
\vspace{0.6cm}

\noindent\large{\textbf{Arthur V. Straube\textit{$^{a}$} and Pietro Tierno$^{\ast}$\textit{$^{b,c}$}}}\vspace{0.5cm}


\vspace{0.6cm}

\noindent \normalsize{We study experimentally and theoretically
the interactions
between paramagnetic particles
dispersed in water and driven above
the surface of a stripe patterned magnetic
garnet film. An external rotating magnetic field modulates
the stray field of the garnet film and generates a translating potential landscape
which induces directed particle motion.
By varying the ellipticity of the rotating field, we
tune the inter-particle interactions
from net attractive to net repulsive.
For attractive interactions, we show that pairs of particles can approach each other and form
stable doublets which afterwards travel along the modulated landscape at a constant mean speed.
We measure the strength of the attractive force
between the moving particles and propose an analytically tractable model
that explains the observations and is in quantitative agreement with experiment.}
\vspace{0.5cm}
 \end{@twocolumnfalse}
  ]

\footnotetext{\textit{$^{a}$~Department of Physics, Humboldt University of
Berlin - Newtonstr. 15, D-12489 Berlin, Germany. E-mail: straube@physik.hu-berlin.de}}
\footnotetext{\textit{$^{c}$~Departament de Estructura i Constituents
de la Mat$\grave{e}$ria, Universitat de Barcelona - Av. Diagonal 647, 08028, Barcelona, Spain. E-mail: ptierno@ub.edu}}
\footnotetext{\textit{$^{d}$~Institute of Nanoscience and
Nanotechnology, IN$^2$UB, Barcelona, Spain}}

\footnotetext{ \ddag~To the memory of Dmitry V. Lyubimov, who shared his exceptional expertise in the methods of averaging and multiple scales}

\section{Introduction}
The transport of particles due to a
ratchet mechanism~\cite{Smoluchowski1912}
is a general phenomenon
arising in many branches
of physics,
and biology.~\cite{Julicher1997,Reimann2002,Hanggi2009}
Ratchet effects are found in Abrikosov vortices,~\cite{Villegas2003}
and Josephson vortices in superconductors,~\cite{Majer2003}
electrons in semiconductor heterostructures,~\cite{Linke1999}
cold atoms,~\cite{Robilliard1999} ferrofluids~\cite{Engel2003} and granular materials~\cite{Meer2004}
to name a few examples.
In biological systems, ratchet effects
are also found in molecular motors
such as myosin~\cite{Cordova1992,Gebhardt2006,Cross2006} or actin.~\cite{Peskin1993,Pantaloni2001,Mogilner2003}

Single particles, molecules or proteins, when placed in an asymmetric potential
will undergo a net transport under non-equilibrium fluctuations.
However, when considering an ensemble of
interacting species,
the system transport properties
are often dictated by a delicate balance between the
particle interactions and the rectification process
above the asymmetric potential.
Unlike molecular machines, or quasi-particles in quantum
systems,
colloidal particles are characterized by experimentally accessible time and length
scales,
and these features promote their use as
a model system to investigate the emergence of novel
ratchet effects.~\cite{Rousselet1994,Faucheux1995,Marquet2002,Lee2005,Tierno2008}
In addition, in colloidal systems
forces and potentials
between the individual particles
can be directly measured
via particle tracking techniques.~\cite{Crocker1994,Kepler1994}

When colloidal particles can be polarized, like paramagnetic colloids,
external fields can be used to induce dipolar
interactions, and assemble these particles into compact structures
such us doublets,~\cite{Calderon2002} chains~\cite{Biswal2004} or clusters.~\cite{Tierno2007}
Magnetic substrates
with features on the colloidal length scale,
have been recently used to
induce directed ratchet transport
of paramagnetic colloidal particles.~\cite{Yellen2005,Tierno2009}
However, most of the recent works concerned the transport of magneticcolloidal particles
focused mainly on the dynamic properties of individual
particles or collective
ensembles, but not on measuring the interaction
forces between the transported particles.
On the other hand,
theoretical works that studied interacting
pairs of particles exhibiting a ratchet-like transport
showed the richness of the physical system.~\cite{Romero2004,Heinsalu2008,Speer2012}

In addition, the use of magnetic fields gives the
freedom to induce attractive or repulsive interactions via dipolar forces.
Thus, the competition between dipolar forces,
\begin{figure*}[!tb]
\begin{center}
\includegraphics[width=0.8\textwidth,keepaspectratio]{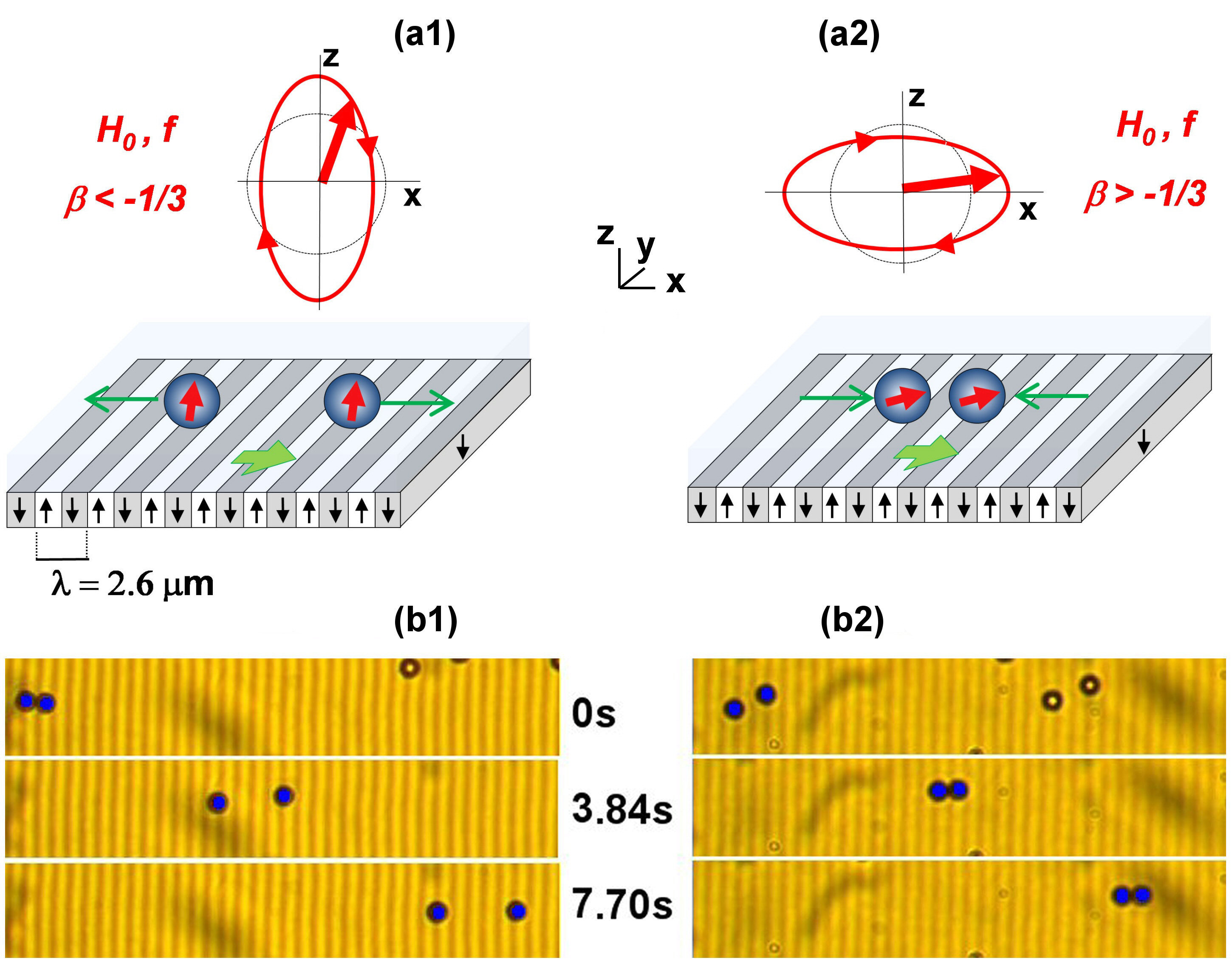}
\caption{(a1,a2)
Schematic illustrations of a pair of paramagnetic
particles transported above the ferrite garnet film (FGF).
The particles display either repulsive (a1) or attractive (a2) interactions
induced by a rotating magnetic field with elliptic polarization.
The field is characterized by a frequency $f$, amplitude $H_0$ and ellipticity parameter $\beta<\beta_{\rm c}$ (a1) or
$\beta>\beta_{\rm c}$ (a2); for particles having no relative displacement along the stripes, $\beta_c=-1/3$.
(b1,b2) Series of optical microscope images at consequent instant showing a
pair of particles (highlighted in blue)
driven above the FGF
and subjected to a magnetic field with $f=10 \, {\rm Hz}$, $H_0=730 \, {\rm A/m}$
and $\beta=-0.6$ (b1), $\beta=0.6$ (b2).}
\label{fig1}
\end{center}
\end{figure*}
which align the particles,
and the substrate field which transports them,
could give rise to novel colloidal structures
and dynamics phases.~\cite{Reichhardt2005,Libal2006,Tierno2012,Jaeger2012}

In this article, we present a detailed study
of the interactions between pairs of
paramagnetic particles driven in
a periodic potential via a deterministic ratchet effect.
The latter is realized by externally modulating the
magnetic stray field generated at the surface of a
ferrite garnet film (FGF).
The modulation corresponding to the rotation of the field breaks the symmetry and induces
a net particle transport above the FGF.
The elliptic polarization of the rotating field is used to tune the inter-particle interactions from net
attractive to net repulsive effects.
The experimental situations
considered are schematically depicted in Fig.~\ref{fig1}(a1,a2).
When the ellipticity of the field
is such that
repulsive interactions dominate (a1),
the paramagnetic colloidal particles either stay disperse,
or couple into oscillating pairs which move above the film.
In the opposite situation, when the
field ellipticity forces the particles to attract each other (a2),
moving particles approach
till forming stable doublets. Afterwards, such doublets
propel above the FGF at a constant mean speed.
We apply a theoretical model that accounts for magnetic
dipolar interactions between the particles driven across the stripes. By integrating out the fast oscillatory
motion caused by the temporal modulation, we put forward an analytically tractable model describing the
dynamics at slow time scales. The theoretical predictions drawn from this model explain
the pair interactions and are in good quantitative agreement with experiment.

\section{Experimental system}

In the experiments, we use a monodisperse suspension of
paramagnetic colloidal particles (Dynabeads M-270, Dynal) with radius $a=1.4 \, {\rm \mu m}$
and magnetic volume susceptibility $\chi \sim 1$.\cite{note1}
The particles were originally dispersed in purified water
at a concentration of $\sim 2 \times 10^{9}$~beads/ml.
We dilute the stock solution with high deionized water
(MilliQ system, $18.2 \, {\rm M \Omega \,\, cm}$)
up to a concentration of $\sim 3 \times 10^{9}$~beads/ml
and deposit a drop of it on top of
the ferromagnetic domains of
an uniaxial ferrite garnet film (FGF).
The FGF film was grown by dipping liquid phase epithaxy
on a gadolinium gallium garnet (GGG)
substrate.~\cite{Helseth2005}
The FGF was characterized by
a series of parallel stripe domains
with opposite magnetization and
spatial periodicity $\lambda = 2.6 \, {\rm \mu m}$,
which is twice the domain
width, Fig.~\ref{fig1}(a).
Between opposite magnetized domains
there are Bloch walls (BWs),
i.e. are narrow transition regions ($\sim 20 \, {\rm nm}$)
where the magnetization rotates, and thus the stray field
of the film is maximal.

After deposition of the droplet,
it takes the particles few minutes to sediment
above the film and get pinned above the BWs.
To prevent particle
adhesion to the magnetic substrate
due to the strong attraction of the BWs,
the FGF was
coated with a $1 {\rm \mu m}$ thick layer of
a photoresist AZ-1512 (Microchem, Newton, MA)
following a protocol
detailed in a previous work.~\cite{Tierno20122}
The polymer film also reduced the
strong attraction
of the BWs, since the stray field of the FGF decreases exponentially
with the elevation.~\cite{Druyvesteyn1972}

The external rotating magnetic field
elliptically polarized in the $(x,z)$ plane
was provided by using two custom-made Helmholtz
coils perpendicular to each other.
The currents in the coils were supplied by
two independent
bipolar amplifiers (Kepco BOP 20-10M, KEPCO)
controlled
with a wave generator
(TGA1244, TTi).
The coils were assembled on the stage
of an upright optical microscope (Eclipse N{\it i}, Nikon)
which was equipped with a $100 \times$ $1.3$ NA
oil immersion objective.
The particle dynamics
was recorded with a
CCD camera (Balser Scout scA640-74fc)
which enabled to grab
videoclips in B/W up to $75$ frames per seconds.
A total field of view of $145 \times 109 \,{\rm \mu m^2}$  was
obtained by adding to the microscope optics
a TV adapter with a lens having a magnification $0.45 \times$.
We measured the positions of the colloidal particles
using a commercial frame-grabbing
software {\it Streampix} (Norpix) and analyzed
the videos with particle tracking routines.~\cite{Crocker1996}
\section{Individual particle dynamics}

Before considering the interactions between particles,
we discuss here the transport mechanism of an individual one
above the FGF.

\begin{figure}[!tb]
\begin{center}
\includegraphics[width=\columnwidth]{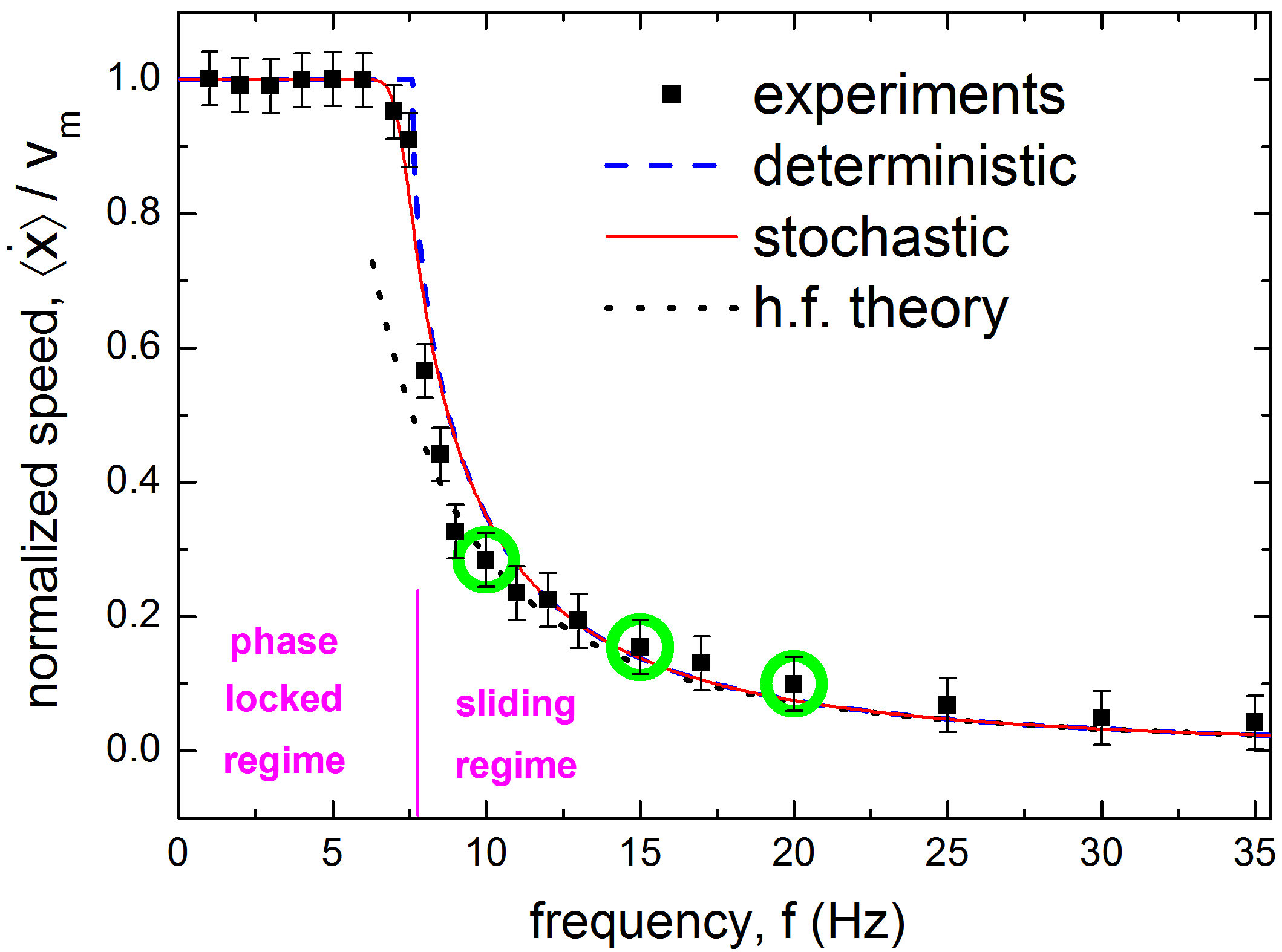}
\end{center}
\caption{Mean speed $\langle \dot{x} \rangle$
of a single particle normalized by $v_{\rm m}=\lambda f$
as a function of frequency $f$ for the case of circular polarization, $\beta = 0$. The deterministic
(dashed line) and stochastic (solid line) theoretical predictions, as in Eqs. (5) and (6), respectively, are fitted
against the experimental data (filled squares). The dotted line corresponds to the high frequency (h. f.)
theory, Eq.~\ref{speed-hi-freq} in the text. Green circles indicate the
cuts at frequencies $f=10$, $15$ and $20 \; {\rm Hz}$ further analyzed in Fig.~\ref{fig3}. }
\label{fig2}
\end{figure}

A paramagnetic particle of radius $a$ and volume $V=(4/3)\pi a^3$,
subjected to an external field ${\mathbf H}$
acquires a dipole moment ${\mathbf m}=V \chi {\mathbf H}$, with $\chi$ being the effective
volume susceptibility of particle.
The energy of interaction of the induced dipole with the magnetic field ${\mathbf B}$ is
$U_{\rm s}=-{\mathbf m} \cdot {\mathbf B}$. Assuming low fields and using the
linear relation ${\mathbf B}=\mu_{\rm s} {\mathbf H}$, where $\mu_{\rm s}$ the permeability of the solvent,
the energy becomes $U_{\rm s}=-V\chi \mu_{\rm s} {\mathbf H}^2$.

The total field above the FGF is given by a superposition $\mathbf{H}=\mathbf{H}^{\rm sub}+\mathbf{H}^{\rm ext}$
of the stray field of the substrate, $\mathbf{H}^{\rm sub}$, and the external field, $\mathbf{H}^{\rm ext}$.
The external field with elliptic polarization has the form:
\begin{equation}
{\mathbf H}^{\rm ext}=(H_{0x} \cos (2\pi f t), 0, -H_{0z} \sin (2\pi f t)),
\label{Hext-dim}
\end{equation}
where $f$ is the frequency. The amplitude of modulation $H_0$ and the
ellipticity parameter $\beta \in [-1,1]$ are defined as:~\cite{Lacis1997}
\begin{equation}
H_0=\sqrt{\frac{H_{0x}^2+H_{0z}^2}{2}}, \quad \beta=\frac{H_{0x}^2-H_{0z}^2}{H_{0x}^2+H_{0z}^2}\,,
\end{equation}
such that $\beta=0$ corresponds to the case of circular polarization.
In all the experiments, we keep $H_0$ fixed, and change the
driving frequency and the ellipticity of the applied field.

The general expression for $\mathbf{H}^{\rm sub}$ can be obtained using the conformal
mapping technique.\cite{Tierno20072, Straube2013} At a moderate modulation, $H_0 \ll M_{\rm s}$, and at a particle
elevation $z\simeq \lambda$, as in our experimental conditions, the expression for the stray field becomes
independent of the form of modulation and can be simplified to:\cite{Straube2013}
\begin{equation}
\mathbf{H}^{\rm sub}=\frac{4 M_{\rm s}}{\pi} {\rm e}^{-2\pi z/\lambda}\left(\cos\frac{2\pi x}{\lambda},0,-\sin\frac{2\pi x}{\lambda}\right),
\label{Hsub1}
\end{equation}
where $M_{\rm s}$ denotes the film saturation magnetization.

The overdamped dynamics of a single particle in the global field ${\mathbf H}$ above
the FGF can be described as the motion in the potential $U_{\rm s}=-V\chi \mu_{\rm s} {\mathbf H}^2$ taken at a fixed
elevation (see Eq.~\ref{eq-Us} in Appendix A), within the framework of the Langevin equation, %
\begin{equation}
\zeta \dot x = -\frac{\partial U_{\rm s}(x,t)}{\partial x} + \sqrt{2 k_{\rm B} T \zeta} \,\xi(t)\,,
\label{LEx-single}
\end{equation}
where $\zeta$ is the viscous friction coefficient, $k_{\rm B}T$ is the thermal energy,
and the stochastic force modeled via the Gaussian white noise with zero mean, $\left<\xi(t)\right>=0$, and the
autocorrelation $\left<\xi(t)\xi(t')\right>=\delta(t-t')$. This model admits a simple interpretation, in
particular we quantify transport by analyzing the averaged speed of the particle.\\

\subsection{Transport in a circularly polarized field, $\beta=0$}

In the case of circular polarization, $\beta=0$, the potential
can be approximated as a traveling harmonic wave,~\cite{Straube2013} $U_{\rm s}(x,t)\propto \cos(2\pi(x/\lambda-v_{\rm m} t))$.
This expression describes a spatially periodic landscape with the period $\lambda$ and minima at
the positions $x_{\rm min}(t) = n \lambda + v_{\rm m} t$ ($n = 0,1,2, \dots$),
which continuously translate with time with a constant speed $v_{\rm m} = \lambda f$ along the $x$ axis.
Further, we proceed to rescaled variables by measuring the length, time, magnetic field, and energy in the units of $\lambda$,
$\zeta\lambda^2/U_0$, $M_{\rm s}$, $U_0$, respectively. We choose the energy unit to be
the characteristic energy of the interaction of an induced dipole with the field generated by the
FGF, $U_0=V\chi \mu_{\rm s} M_{\rm s}^2$.

In these units, the averaged speed of the particle can be calculated as:~\cite{Straube2013}
\begin{equation}
\frac{\left<\dot x\right>_{\beta=0}}{v_{\rm m}} = \left\{
\begin{array}{ll}
1, & {\rm if} \;\; \tilde{f} < \tilde{f}_{\rm c}(0)\,, \\
1 - \sqrt{1-\tilde{f}_{\rm c}^2(0)/\tilde{f}^2}\,, & {\rm if} \;\; \tilde{f} > \tilde{f}_{\rm c}(0)\,, \\
\end{array}
 \right.
\label{speed-det-beta=0}
\end{equation}
without thermal fluctuations and,
\begin{equation}
\frac{\left<\dot x\right>_{\beta=0}}{v_{\rm m}} = 1-\frac{\sinh(\pi D)}{\pi D \,|I_{iD}(D_{\rm c})|^2}\,
\label{speed-stoch-beta=0}
\end{equation}
with thermal fluctuations. Here, we have introduced three parameters,
\begin{equation}
h_0=\frac{H_0}{M_{\rm s}}, \quad \tilde{f}=\frac{f \zeta \lambda^2}{U_0}, \quad
\sigma=\frac{k_{\rm B} T}{U_0},
\end{equation}
which are, in order, the dimensionless amplitude, frequency, and strength of thermal fluctuations. Then,
$\tilde f_{\rm c}(0) = 16 h_0 {\rm e}^{-2\pi z}$ is the critical frequency at $\beta=0$, $D=\tilde f/(2\pi \sigma)$,
$D_{\rm c}=\tilde f_{\rm c}(0)/(2\pi \sigma)$, and $I_{i\nu}(x)$ is the modified Bessel
function of the first kind of an imaginary order.

\begin{figure}[!tb]
\begin{center}
\includegraphics[width=\columnwidth]{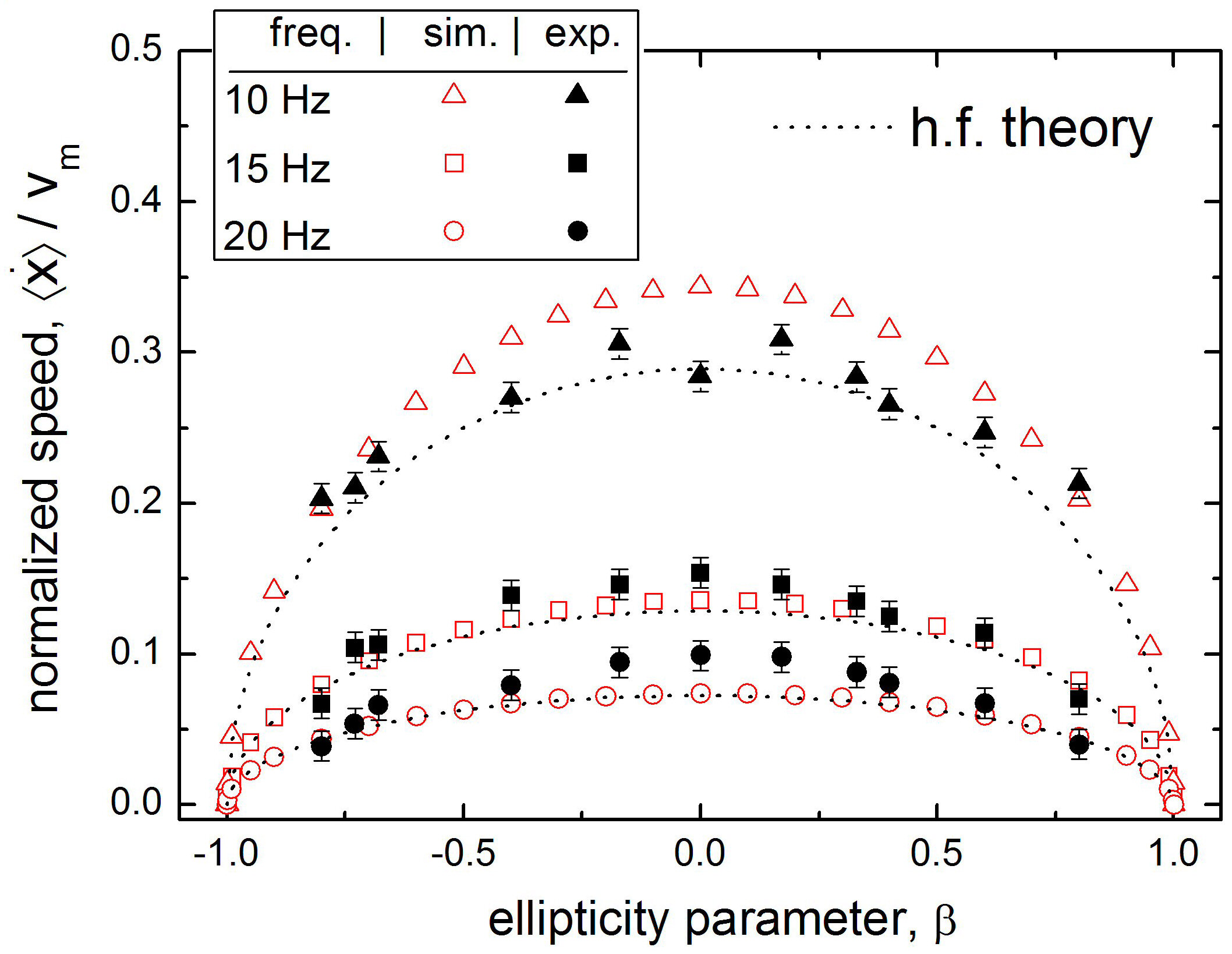}
\end{center}
\caption{The normalized mean speed, $\left<\dot x\right>/v_{\rm m}$, as a function of the
ellipticity parameter, $\beta$, at three different frequencies, as shown by green circles in Fig.~\ref{fig2}.
The experimental data (filled markers) are plotted against the
predictions of numerical simulations (open markers), Eq.~\ref{LEx-single}, and of
the h.f. theory (dotted lines), Eq.~\ref{speed-hi-freq}.  }
\label{fig3}
\end{figure}
From Eqs.~\ref{speed-det-beta=0} and~\ref{speed-stoch-beta=0}
it follows that, increasing the driving frequency,
the system is characterized by two dynamic states
separated by the critical value $\tilde f_{\rm c}$.
This behaviour is also illustrated in Fig.~\ref{fig2},
where we report measurements
of the average speed of a single particle as a function of the driving frequency.
The paramagnetic particle is driven above a garnet film by a circularly polarized ($\beta=0$)
magnetic field with the amplitude $H_0=730 \, {\rm A/m}$.
At low frequencies, the particle is trapped in the minima of the translating potential,
and moves with the maximal speed, $v_{\rm m}$.
Beyond a critical frequency of $f_{\rm c} \approx 7.6 \, {\rm Hz}$,
the particle starts to lose its synchronization with the moving landscape
entering into a ``sliding'' regime, where it decreases its average speed.
Fig.~\ref{fig2} also shows that
thermal fluctuations smooth the transition from the phase-locked dynamics to the
sliding motion near the critical point.
By fixing the particle elevation
above the film to $z=0.923$ (in the units of $\lambda$),
we estimated the dimensionless amplitude $h_0\approx 0.1457$ and noise strength of $\sigma \approx 2 \times 10^{-5}$.

\subsection{Transport in an elliptically polarized field, $\beta \ne 0$}

The transition between the locked and sliding phases illustrated in Fig.~\ref{fig2} occurs also
for different values of $\beta$, i.e. when the modulation has elliptic polarization.
In particular, the critical frequency $\tilde f_{\rm c}$ depends on $\beta$, and we find
that it shifts to lower frequencies, $\tilde f_{\rm c}(\beta)<\tilde f_{\rm c}(0)$.
To gain insight into the sliding dynamics of a single particle
at $\beta \ne 0$, we perform the time averaging of Eq.~\ref{LEx-single}
taken in the deterministic limit, $\sigma=0$.
The latter is justified by the fact that, as shown in Fig.~\ref{fig2},
thermal fluctuations play a negligible role away from the
critical frequency.
As a result, the mean speed of a single particle is given by:
\begin{equation}
\frac{v_0(\beta)}{v_{\rm m}}=\frac{\left<\dot x\right>_{\rm hf}}{v_{\rm m}}
= \frac{1}{2} \left( \frac{16 h_0}{\tilde f} \right)^2 {\rm e}^{-4 \pi z} \sqrt{1-\beta^2}
\quad (\tilde f \gg \tilde f_c)\,
\label{speed-hi-freq}
\end{equation}
valid for any $\beta$ at high frequencies.
A complete derivation of Eq.~\ref{speed-hi-freq}
is given in Appendix~\ref{sec:AppendixA}.
The accuracy of this prediction can be estimated from Fig.~\ref{fig2}.
Although the h. f. analysis is formally valid in
the high frequency limit, $\tilde f/\tilde f_{\rm c} \gg 1$, we
see that it works well already at $\tilde f/\tilde f_{\rm c}(0) \approx 2$ ($15 \; {\rm Hz}$)
and is still reasonable even at the lower frequency of $10 \; {\rm Hz}$.

\begin{figure}[!tb]
\begin{center}
\includegraphics[width=\columnwidth,keepaspectratio]{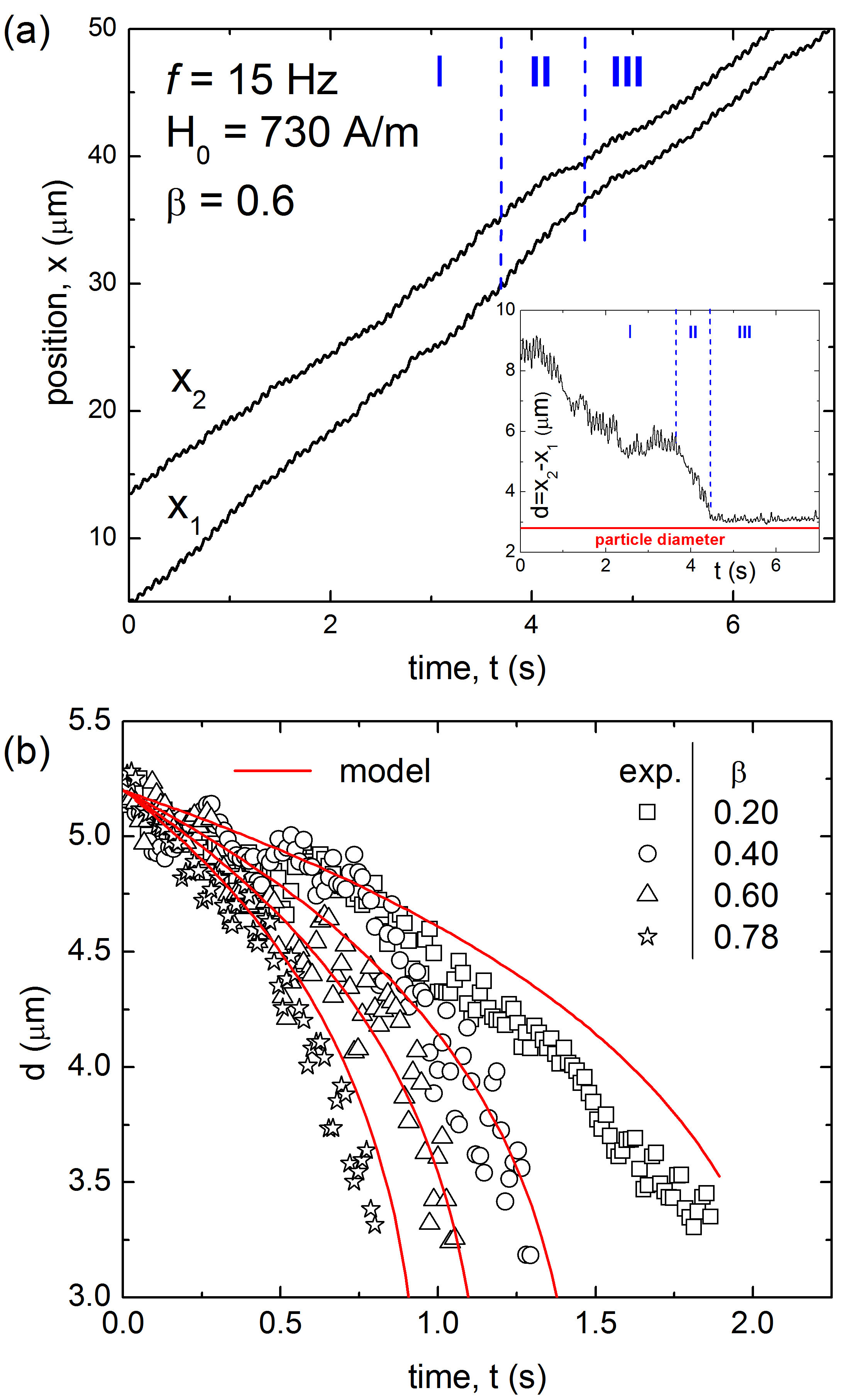}
\end{center}
\caption{(a) Positions $x_1$ and $x_2$ versus time $t$
of two approaching paramagnetic colloidal particles subjected to an external field
with the frequency $f=15 \, {\rm Hz}$, amplitude $H_0 = 730 \, {\rm A/m}$,
and ellipticity $\beta = 0.6$. We distinguish three
regimes: one characterized by a slow approach of the particles (I), a second characterized
by dipolar attraction and leading to the
doublet formation (II), and finally the last where particle motion occur
in form of a doublet (III). Inset shows
the separation distance $d=|x_2-x_1|$ versus time, $t$.
(b) Separation distance $d$ versus time $t$ in regime II plotted at different $\beta$.
Scattered points are experimental data, solid red lines
are fits following the theoretical model, see Eq.~\ref{d(t)} in the text.   }
\label{fig4}
\end{figure}

In Fig.~\ref{fig3} we show the impact of the
ellipticity of the field, $\beta$, on the average speed $\langle \dot{x} \rangle$
of a single particle and at three different driving frequencies.
For circularly polarized field ($\beta = 0$), $\langle \dot{x} \rangle$
is maximum for all frequencies, and it decreases as $\beta\ne 0$,
in a symmetric way with respect to the positive and negative values of $\beta$ according to the
root law $\left<\dot x\right>/v_{\rm m} \propto \sqrt{1-\beta^2}$.
The experimental results are in good agreement with the predictions
from Brownian dynamics simulation using Eq.~\ref{LEx-single} and the h. f. theory, Eq.~\ref{speed-hi-freq},
as described in Appendix A.
Fig.~\ref{fig3} also shows that the h. f.
approximation well represent the dependence of $\left<\dot x\right>$ on $\beta$.

\section{Interacting particles}

Increasing the number of particles,
forces the latter to interact via magnetic dipolar interactions,
see Appendix~\ref{sec:AppendixB}
for details. Experimentally,
we observed a different behaviour
depending whatever the particles were moving in the
phase locked or in the sliding regime.
In the first regime, the particles formed
a series of chains equally spaced along the direction of motion ($x$),
and all of them were moving at
same average speed, $v_{\rm m}$.
In this situation, even for large ellipticity,
the particles always keep the difference in their $x$ coordinates constant,
and it was not possible to induce attraction or repulsion,
breaking the robust dynamic pattern.
In contrast, in the sliding regime, each particle was unable to follow the fast dynamics of the translating potential
and it lost the phase-locking with the field at different times.
Since this process did not occur synchronously for all the particles,
the moving colloids showed a certain degree of randomization in their speeds.
As a consequence, between each pair of particles the average distance along $x$ was not always fixed, but
it could increase or decrease depending on the relative speed. Thus,
in the sliding regime, we found that it was possible to
tune the particle interaction by changing $\beta$.

\subsection{Two particles moving one behind another}

To study the effects caused by the dipole-dipole interactions, we first analyze the one-dimensional
situation in which a pair of particles has no relative displacement along
the stripes ($y_1=y_2$), moving one behind the other in the sliding regime.

Fig.~\ref{fig4} shows the time evolution of the positions $x_1$ and $x_2$
of a pair of colloidal particles initially placed at a relative distance
of $d=8.2 \, {\rm \mu m}$, and driven above an FGF by an elliptically polarized magnetic field with
amplitude $H_0=730 \, {\rm A/m}$, frequency $f=15 \, {\rm Hz}$ and ellipticity $\beta=0.6$. As we show below in
this section, this value of $\beta$
corresponds to attracting dipolar interactions.
The displacements shown in Fig.~\ref{fig4}, illustrate the three regimes of motion. In the first one (regime I),
the separation distance is too large to cause an evident effect of attraction,
and the particles slowly approach each other
due to a small difference in their speeds in the sliding regime.
The relative dynamics is
governed by the interplay between thermal fluctuations and the driving potential. Note that the separation distance $d=d(t)$ displays pronounced
oscillations. As explicitly shown in Appendix~\ref{sec:AppendixA}, these oscillations
are caused by the external modulation and occur with the external frequency $f$. When the particles come close enough,
to about $d \simeq 5.2 \, {\rm \mu m}$ in our case, their relative motion speeds up and their distance $d$ rapidly decreases to a minimal
distance dictated by steric interactions (regime II).
After that the particles have formed a stable doublet (regime III) and propel as a whole.
Note, however, that $d(t)$ does not remain equal to exactly the hard-core distance of $2a$.

To address the one-dimensional problem theoretically,
we will apply the h.f. theory
developed in Appendix~\ref{sec:AppendixC}. The interaction of the
%
\begin{figure}[!tb]
\begin{center}
\includegraphics[width=\columnwidth,keepaspectratio]{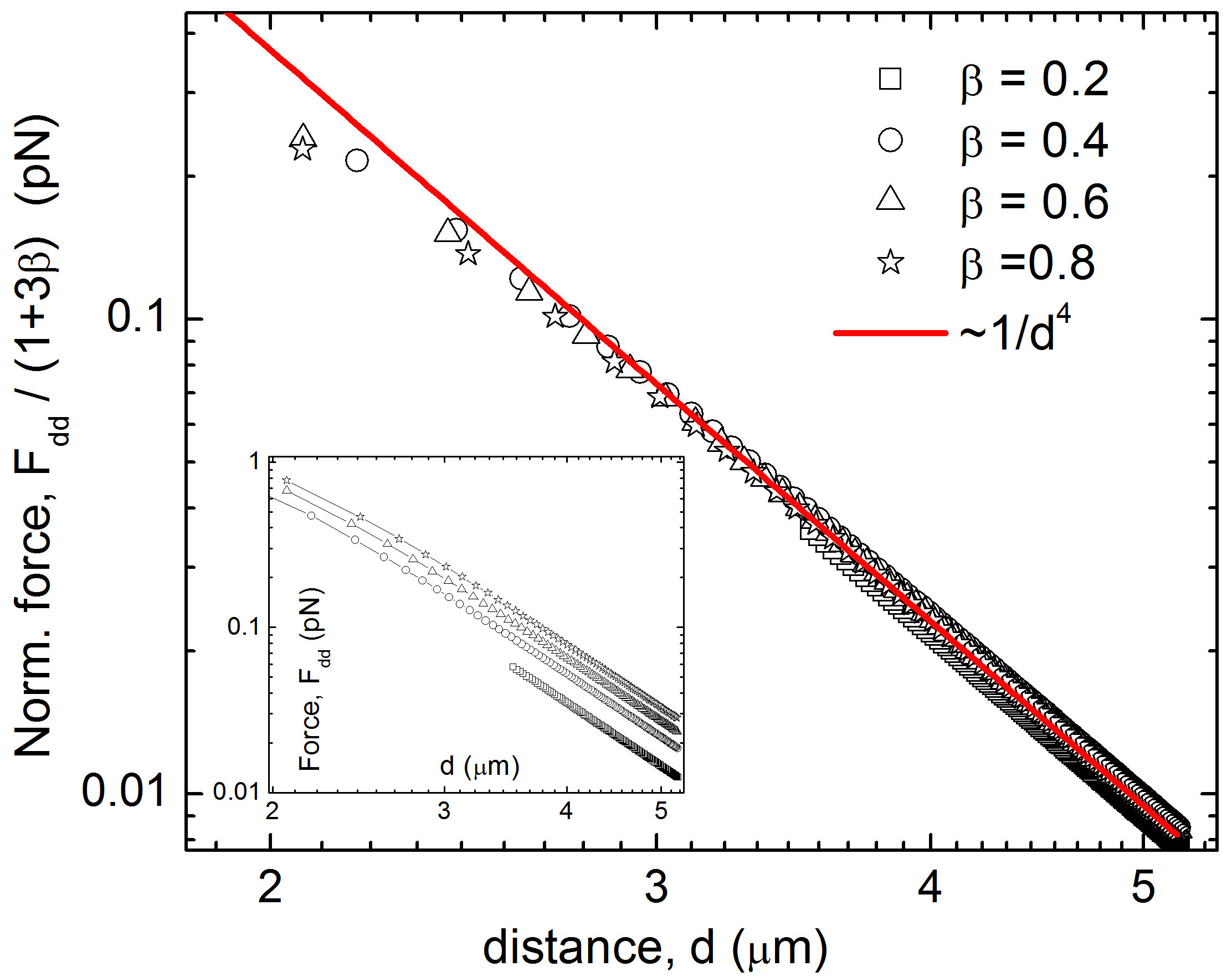}
\end{center}
\caption{Log-log plot of the force $F_{\rm dd}$ between a pair of particles normalized by $(1+3\beta)$
and plotted as a function of the separation distance $d$.
Scattered data correspond to the lines fitting the experimental points in Fig.~\ref{fig4}(b), solid red line is
a fit according to Eq.~\ref{dotd-F(d)}, showing the dipolar nature of the interaction.
Inset shows the force versus distance without the normalization for different values of $\beta$.  }
\label{fig5}
\end{figure}
%
two particles with the slowly evolving coordinates ${\mathbf R}_1=(X_1, Y_1)$ and
${\mathbf R}_2=(X_2, Y_2)$ is described by the effective potential
given by Eq.~\ref{Udd-hf} or Eq.~\ref{Udd-hf-theta}. Taking into
account that $Y_1=Y_2$ (or $\vartheta=0$, where $\vartheta$ is the angle between the axis $x$ and the straight line going through the centers of particles) and introducing the distance between
the particles as $d:=|X_{12}|=|X_1-X_2|$, we have $R=d$, $X_{12}^2/R^2=1$. Hence, the
effective interaction potential that describes the slow dynamics of particles simplifies to
\begin{equation}
U_{\rm dd}(d)=-\frac{\alpha h_0^2(1+3\beta)}{2d^3}\,.
\label{Udd-hf-1d}
\end{equation}
Whether the particles attract or repel
depends on the sign of the factor $1+3\beta$. Setting it to zero,
we find that the critical value is
\begin{equation}
\beta_{\rm c}(\vartheta=0)=-\frac{1}{3}\,. \label{beta-c-1d}
\end{equation}
For $\beta < \beta_{\rm c}$ the particles repel each other, while for $\beta > \beta_{\rm c}$
attraction takes place.

The separation distance satisfies the dimensionless equation $\dot d = -2\partial_d U_{\rm dd}=-3\alpha h_0^2(1+3\beta)/d^4$. Rewriting this equation back in the original variables, as before re-scaling, we obtain
\begin{equation}
\zeta \dot d = -\frac{k(1+3\beta)}{d^4}=:F_{\rm dd}(d)\,,
\label{dotd-F(d)}
\end{equation}
where the constant $k=3\mu_s(\chi V H_0)^2/(4\pi)$. Thus, at a given field amplitude, $H_0$, the strength of interactions between a pair
of particles scales with the ellipticity of the field, $\beta$, the susceptibility $\chi$ and size $a$ of particles as $F_{\rm dd} \propto (1+3 \beta) \chi^2 a^6$.

Assuming that at time $t=0$ the particles are initially separated by a distance $d=d_0$, we integrate Eq.~\ref{dotd-F(d)} to find a power law for the separation distance as a function of time:
\begin{equation}
d(t)=\left(d_0^5-\frac{5k(1+3\beta)}{\zeta}t\right)^{1/5}\,, \label{d(t)}
\end{equation}
From Eq.~\ref{d(t)} follows that for $\beta<\beta_c(0)=-1/3$ ($\beta>\beta_{\rm c}(0)$), the separation distance increases (decreases) with time.
During attraction, the particles approach till reaching a minimal distance
$d_{\rm m}$ which for hard spheres is given by, $d_{\rm m}=2a$.
From Eq.~\ref{d(t)} it is possible also to
estimate the time  takes the particles to come into contact, as  $\tau_{\rm c}=\zeta(d_0^5-d_{\rm m}^5)/[5k(1+3\beta)]$.

\begin{figure*}[!tb]
\begin{center}
\includegraphics[width=0.8\textwidth,keepaspectratio]{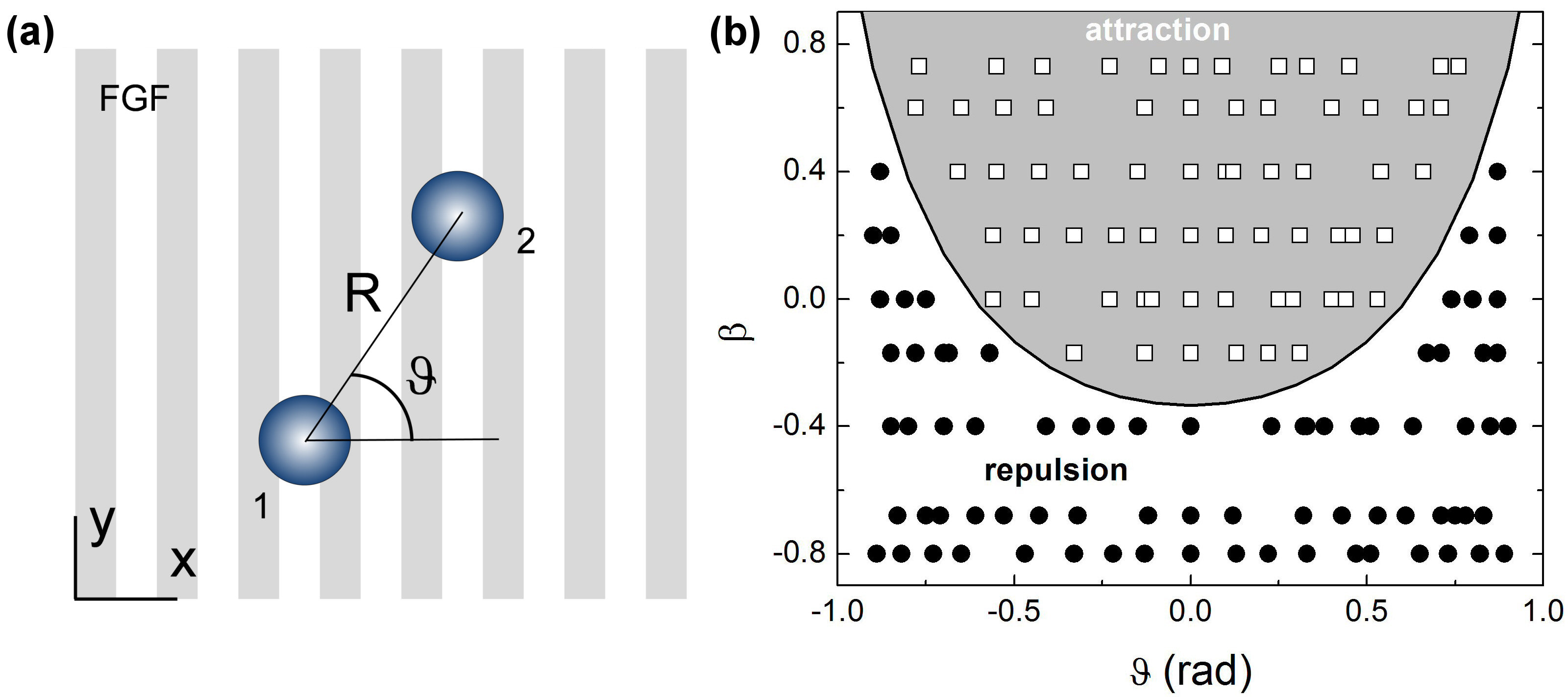}
\end{center}
\caption{(a) Schematic showing a pair of
interacting particles driven above the FGF and having arbitrary positions in the plane $(x,y)$.
(b) Phase diagram in the plane $(\vartheta,\beta)$, showing the regions of attraction and repulsion.
Here, $\vartheta$ denotes the polar angle introduced as shown in panel (a).
Scattered data are experimental points, solid line
is according to Eq.~\ref{beta-c-2d}.}
\label{fig6}
\end{figure*}

In order to directly derive the strength of the dipolar interactions from the experimental data,
we estimated the dependence of the force $F_{\rm dd}$ on the separation distance $d$.
The inset of Fig.~\ref{fig5} shows the dependencies $F_{\rm dd}(d)$ for different $\beta$. The values of the force were computed using the Stokes law, $F_{\rm dd}=\zeta v_{\rm d}$, where the speeds $v_{\rm d}$
were recovered from the solid red curves in Fig.~\ref{fig4}(b) that fit the experimental data. The friction coefficient was drawn from the relation $\zeta = 6\pi \eta a$, where  $\eta=10^{-3} \, {\rm Pa \cdot s}$
is the dynamic viscosity of water. Following Eq.~\ref{dotd-F(d)},
we expect the ratio $F_{\rm dd}/(1+3\beta)=k/d^4$ to be independent
of the field ellipticity, $\beta$. This prediction is validated in Fig.~\ref{fig5},
by plotting the force $F_{\rm dd}$ normalized by $1+3\beta$ as a function of the distance $d$.
We note that all the dependencies for the different values of $\beta$ showed in the inset, collapse into the same curve.
Furthermore, from the regression we obtain a value of the constant $k\approx 5.91 \,{\rm pN \; \mu m^4}$,
which is in good agreement with the theoretical prediction $k=3\mu_s(\chi V H_0)^2/(4\pi)\approx 5.93 \,{\rm pN \; \mu m^4}$, evaluated based on the experimental parameters, taking into account the uncertainty related to
the exact value of $\chi$.\cite{note1}
The magnetic permeability of the solvent was estimated as the permeability of free space.

We note that Eqs.~\ref{dotd-F(d)} and \ref{d(t)} present purely deterministic predictions for the dipolar force and the separation distance.
Similarly to the situation of a single particle, as e.g., in Fig.~\ref{fig2}, thermal fluctuations are expected to slightly
slow down the deterministic dynamics
in regime II, as in Eq.~\ref{d(t)}. As confirmed by Brownian dynamics simulations,
results not shown here,
the thermal noise indeed effectively weakens the attractive forces shortly before the
particles come into contact, thus slightly increasing the time of approach of the particles in regime II.
This tendency can be also seen from Fig.~\ref{fig5}, where the
experimental data start to undershoot the deterministic predictions at small $d$,
close to the smallest particle distance.

\subsection{Particles with arbitrary positions}\label{ssec:arb-posit}

We now consider the general situation in which a pair of particles
have arbitrary positions in the $(x,y)$ plane, and using the
h. f. theory.
First, we mention the motion of the center of mass of the
two particles. The equation of motion for the center of mass,
${\mathbf Q}=({\mathbf R}_1+{\mathbf R}_2)/2$, can be deduced
from Eq.~\ref{eqs-XlY1} in Appendix C. The center of mass moves strictly across the stripes with
the constant speed of a single particle, and there is no
displacement along the stripes, $\dot{\mathbf Q}=(\dot Q,0)=v_0 \hat{\mathbf e}_x$, irrespective of the positions of the particles in the plane $(x,y)$.

Then, we analyze the relative motion of particles. Instead of the
Cartesian coordinates ${\mathbf R}_{12}=(X_1-X_2, Y_1-Y_2)$, it is
convenient to proceed to the polar coordinates $(R,\vartheta)$
introduced such that ${\mathbf R}_{12}=R (\cos\vartheta, \sin\vartheta)$,
where $R=\sqrt{(X_1-X_2)^2+(Y_1-Y_2)^2}$ is the distance between the
particles, see Fig.~\ref{fig6}(a). After the transformation, the equations of
motion $\dot R=-2\partial_R U_{\rm dd}(R,\vartheta)$ and
$R^{2}\dot \vartheta =-2\partial_{\vartheta} U_{\rm dd}(R,\vartheta)$
with $U_{\rm dd}(R,\vartheta)$ given by Eq.~\ref{Udd-hf-theta}, result in:
\begin{eqnarray}
\dot R & = & \frac{3\alpha h_0^2}{R^4}\left[2 - 3(1+\beta) \cos^2\vartheta \right], \label{eq-R}\\
\dot \vartheta & = & -\frac{3\alpha h_0^2 (1+\beta)}{R^5}\sin2\vartheta\,.\label{eq-theta}
\end{eqnarray}
By setting $\dot R =0$ in Eq.~\ref{eq-R} we consider
the marginal case that separates the situations of repulsion, $\dot R >0$,
and attraction, $\dot R <0$. This condition gives us the critical value of
the ellipticity parameter,
\begin{equation}
\beta_{\rm c}(\vartheta)=-1+\frac{2}{3 \cos^2\vartheta}\,, \label{beta-c-2d}
\end{equation}
generalized for arbitrary values of $\vartheta$. Again, the
condition $\beta < \beta_{\rm c}$ corresponds to repulsion, while
the opposite case $\beta > \beta_{\rm c}$ is responsible for attraction.
In the partial case of the particles moving along the $x$ direction, $\vartheta=0$,
Eq.~\ref{beta-c-2d} predicts $\beta_{\rm c}(0)=-1/3$, in agreement
with the earlier considered case, see Eq.~\ref{beta-c-1d}. The
opposite partial case of particles traveling across the stripes side by
side, $\vartheta=\pi/2$, is always repulsive, which is seen
from Eq.~\ref{eq-R}, since $\dot R >0$. A repulsion-attraction diagram, which demonstrates agreement between the theory and experiment,
is shown in Fig.~\ref{fig6}(b).

We note that this analysis implies that the angle $\vartheta$ is
constant and refers not only to a given position but also to a given instant
of time. However, the polar angle $\vartheta$ generally evolves in time. As
follows from Eq.~\ref{eq-theta}, it admits two fixed points,
$\vartheta_0^{(1)}=0, \pi$ and $\vartheta_0^{(2)}=\pm \pi/2$.
The first one, when the particles move one behind another across the stripes, is stable.
The second one, when the particles travel across the stripes side by side and attract or repel along the stripes,
is unstable. The evolution of the angle is determined by the sign
of $\sin 2\vartheta$ and we conclude that independent of the
ellipticity $\beta$, the particles evolve towards the stable
state with $\vartheta=0, \pi$. In other words, the particles
tend to reorient such that the straight line through the centers of particles aligns along the $x$ axis.

\section{Conclusions}

In this article, we studied both experimentally and
theoretically the dynamics of interacting paramagnetic
colloidal particle magnetically driven above
a stripe patterned garnet film. We show that
attractive dipolar interactions between
propagating particles become important for distances
lower than $d_0 \sim 6 {\rm \mu m}$
for the used field strength of $H_0=730 \, {\rm A/m}$, although this distance can be tuned by
changing the amplitude of the applied field $H_0$.
When particles approach closer than $d_0$,
they form stable doublets
which move at a constant mean speed along the modulated landscape.

The suggested theoretical model, which describes the slow dynamics of interacting particles averaged over the fast oscillatory time scale, is analytically tractable. It captures the experimental results quantitatively well. In particular, we gain an insight into the details underlying the interaction, by outlining an effective interaction potential.
These findings can be used to extend the model towards more complicated situations,
involving a large number of particles or binary mixtures driven above a garnet film. On the other hand,
the application of a similar approach is potentially promising
for studying the transport of interacting particles in
other systems using magnetic structure substrates.~\cite{Yellen2007,Yellen2010,Yellen2011,Gunnarsson2005,Ehresmann2011}

The possibility to tune the sign of the inter-particle interactions and
their relative strength in transport at small scales
has potential applications in microfluidics and lab-on-a-chip
systems. In particular, it can be used
to pick up and capture a microscopic cargo
between attractive particles, transport
and finally release it at a prescribed location by switching the attractive interaction to become repulsive.

Furthermore, the use of attractive interactions between the moving particles
can be used to generate longer chains
traveling along the modulated landscapes, as shown
for smaller particles.~\cite{Tierno2012}
These chains can serve as a model to study
fluctuations in driven Brownian worms,\cite{Toussaint2004} or novel ratchet effects
arising from condensed particle trains.\cite{Dzubiella2002,Reichhardt2006,Pototsky2010}

\section*{Appendix}
\appendix

\section{Slow dynamics of a single particle}\label{sec:AppendixA}

At high frequencies, different times scales naturally present
in the system become well separated and admit the possibility to
reduce the complexity by effectively decoupling the fast and slow motions.
\cite{Straube2006} The ``fast'' dynamics is associated with the external
driving with the characteristic time scale $\tau_{\rm f}=1/f$. The ``slow''
motion, such as, propulsion of a single particle across the
stripes in our system, is the ``net'' or mean (time-averaged) response
of the system at time scales $t \gg \tau_{\rm f}$.

We now consider the overdamped motion of a single particle in the field ${\mathbf H}$ above the substrate, which is described by the dimensionless potential
\begin{equation}
U_{\rm s}(x,t)=-\frac{8h_0}{\pi}{\rm e}^{-2\pi z}\left[u_1\cos(2\pi x)+u_2\sin(2\pi x)\right] \label{eq-Us}
\end{equation}
with $u_1(\beta,t) = \sqrt{1+\beta}\cos(2\pi \tilde f t)$ and $u_2(\beta,t)=\sqrt{1-\beta}\sin(2\pi \tilde f t)$.
To obtain the description for the slow motion of the particle, we have to perform
a time averaging of Eq.~\ref{LEx-single}
without thermal noise
\begin{eqnarray}
\dot x(t) & = & - \partial_x U_{\rm s}(x,t) =F_{\rm s}(x,t)\,, \label{dotx-eq-single-det} \\
F_{\rm s}(x,t) & = & -16 h_0 {\rm e}^{-2\pi z}\left[u_1\sin(2\pi x)
- u_2\cos(2\pi x) \right]. \quad \label{Fs(x,t)}
\end{eqnarray}
The problem is considered deterministic, $\sigma=0$,
because, as explained in the main text, thermal fluctuations are negligible
for high-frequencies, $\tilde f \gg \tilde f_{\rm c}$.
Following the method of averaging,\cite{Nayfeh1981,Piet2013,Piet20132}
we present the solution as a superposition:
\begin{equation}
x(t)=X(t)+\delta x(t), \quad
\delta x(t)=\tilde x {\rm e}^{2\pi i \tilde f t} + \tilde x^{\ast} {\rm e}^{-2\pi i \tilde f t}\,,
\label{ansatz-x(t)-hf}
\end{equation}
where $X(t)$ and $\delta x(t)$
describe the slow (time-averaged over the period $1/\tilde{f}$ of modulation)
coordinate and its fast (time-periodic) counterpart oscillating with the frequency
$\tilde f$, respectively. The quickly evolving contribution $\delta x(t)$, which has
to be considered small compared to $X(t)$, is then represented via the complex
amplitude $\tilde x$ and its complex conjugated pair $\tilde x^{\ast}$, as in
Eq.~\ref{ansatz-x(t)-hf}. The complex amplitudes do not explicitly
depend on the fast time $\tilde f t$. We note that it is convenient to
use exponential representation of the functions $\cos(2\pi\tilde f t)$ and
$\sin(2\pi\tilde f t)$. The spatially dependent functions $\cos(2\pi x)$
and $\sin(2\pi x)$ are expanded using the smallness of $\delta x$,
according to $g(x)=g(X+\delta x) \approx g(X)+\partial_x g(X) \delta x$.

Substituting the ansatz \ref{ansatz-x(t)-hf} into Eq.~\ref{dotx-eq-single-det},
using the described representations, and retaining the leading terms, we find for the
complex amplitude:
\begin{equation}
\tilde x(X) = \frac{4 h_0}{\pi \tilde f} {\rm e}^{-2\pi z}\left[ i \sqrt{1+\beta}\sin(2\pi X)
- \sqrt{1-\beta}\cos(2\pi X) \right].
\label{x-puls}
\end{equation}
To obtain the equation for $X$,
we perform the time-averaging of Eq.~\ref{dotx-eq-single-det}.
We evaluate the time-averaged contributions,
$\overline{\sin(2\pi x)\cos(2\pi \tilde f t)}= \pi (\tilde x^{\ast}+\tilde x) \cos(2\pi X)=-(8h_0/\tilde f){\rm e}^{-2\pi z}\sqrt{1-\beta}\cos^2(2\pi X)$
and $\overline{\cos(2\pi x)\sin(2\pi \tilde f t)}= i\pi (\tilde x^{\ast}-\tilde x) \sin(2\pi X)=-(8h_0/\tilde f){\rm e}^{-2\pi z}\sqrt{1+\beta}\sin^2(2\pi X)$. Here, the
overlines denote the time averaging over the period of modulation, $\overline{\mathcal{F}}=\tilde f\int_0^{1/\tilde f} \mathcal{F}\, dt$, and the combinations $\tilde x^{\ast} +\tilde x=2\,{\rm Re}(\tilde x)$ and $i(\tilde x^{\ast} -\tilde x)=2\,{\rm Im}(\tilde x)$ are evaluated via the real and imaginary parts of Eq.~\ref{x-puls}. As a result, the time averaged equation takes a simple form
\begin{equation}
v_0(\beta):=\dot X =\left<\dot x\right>_{\rm hf}= \frac{1}{2}\frac{(16 h_0)^2}{\tilde f} {\rm e}^{-4\pi z} \sqrt{1-\beta^2}\,, \label{v0}
\end{equation}
which, being written relative to the maximal
speed, $v_{\rm m}=\tilde f$, gives Eq.~\ref{speed-hi-freq}.

Because the equation for the slow dynamics of a single particle
is independent of $X$ and $t$, it means that the particle moves
on the average with a constant speed. Therefore, expression \ref{v0}
is interpreted as the mean speed in the sliding regime, valid at high
frequencies and at all $\beta$. As follows from Eq.~\ref{v0}, the time
averaged motion of a single particle is equivalent to the motion in the mean potential
\begin{equation}
U_{\rm s}(X)=-v_0 X\,. \label{Us(X)}
\end{equation}
It should be noted that time averaging directly the
potential in favor of the equations of motion, can lead to misleading results.
For instance, performing the averaging of Eq.~\ref{eq-Us} does not lead to Eq.~\ref{Us(X)}
but results in identically vanishing $U_{\rm s}(X)$, which incorrectly predicts no motion.

\section{Magnetic dipolar interactions}\label{sec:AppendixB}

In a suspension of magnetic dipoles, each dipole interacts with the fields produced by all other dipoles.
Induced dipole $l$ with the magnetic moment ${\mathbf m}_l = V \chi {\mathbf H}_l$
interacts with the field ${\mathbf B}_{l'}=\mu_s {\mathbf H}_{l'}$ generated by particle $l'$, leading
to the dipolar energy $U_{\rm dd}=-{\mathbf m}_l \cdot {\mathbf B}_{l'} = -{\mathbf m}_{l'} \cdot {\mathbf B}_l$.
Thus, for a system of dipoles with the coordinates ${\mathbf r}_l$
the total energy can be written as:
\begin{equation}
U=\sum_l U_{\rm s}({\mathbf r}_l,t) +\frac{1}{2} \sum_{l}\sum_{l'\ne l} U_{\rm dd}({\mathbf r}_{ll'},t)\,. \label{U-tot-dim}
\end{equation}
Here, the first contribution stands for the interaction of each single dipole with the nonuniform magnetic field above the FGF and the second term describes the dipolar interactions with the pairwise potential
\begin{equation}
U_{\rm dd}({\mathbf r}_{ll'},t) = -\frac{\mu_s V^2 \chi^2}{4 \pi} \left[3\,\frac{{\mathbf H}_l\cdot {\mathbf r}_{ll'}\, {\mathbf H}_{l'} \cdot {\mathbf r}_{ll'} }{r_{ll'}^5} - \frac{{\mathbf H}_l \cdot {\mathbf H}_{l'} }{r_{ll'}^3} \right], \label{Udd-dim}
\end{equation}
where ${\mathbf H}_l={\mathbf H}({\mathbf r}_l, t)$, ${\mathbf r}_{ll'}={\mathbf r}_{l}-{\mathbf r}_{l'}$, and $r_{ll'}=|{\mathbf r}_{ll'}|$.
By measuring the lengths in the scale of $\lambda$ and energy in the units of
$U_0=V \chi \mu_s M_{\rm s}^2$ as before and accounting for Eq.~\ref{Udd-dim}, the dimensionless expression
for the total energy,  Eq.~\ref{U-tot-dim}, becomes
\begin{equation}
U = - \sum_l {\mathbf H}^2_l
-\frac{1}{2} \alpha \sum_{l}\sum_{l'\ne l}
\left[3\,\frac{{\mathbf H}_l\cdot {\mathbf r}_{ll'}\, {\mathbf H}_{l'}\cdot {\mathbf r}_{ll'} }{r_{ll'}^5}
- \frac{{\mathbf H}_l \cdot {\mathbf H}_{l'} }{r_{ll'}^3} \right].
\label{U-tot-dimless}
\end{equation}
The dimensionless parameter
\begin{equation}
\alpha =\frac{\chi}{4\pi} \frac{V}{\lambda^3} = \frac{\chi}{3} \left(\frac{a}{\lambda}\right)^3\, \label{alpha}
\end{equation}
determines the strength of dipole-dipole interactions relative to the energy of interaction with the FGF, $U_0$.
For our experimental system, $\alpha\approx 0.027$, if $\chi = 0.53$.

\section{Slow dynamics of two interacting particles}\label{sec:AppendixC}

The interaction potential
between the driven particles taking into account the dipolar interactions
is quite complicated, since it consists of different contributions
resulting from the temporal
modulation, the field of substrate and their interplay,
described by terms of order $\mathcal{O}(h_0^2)$, $\mathcal{O}(h_0{\rm e}^{-2\pi z})$,
and $\mathcal{O}({\rm e}^{-4\pi z})$, respectively. At our experimental
conditions ($h_0 \ll 1$, $z \simeq 1$), the mean drift of particles is due to the
interplay of temporal modulation and the field of substrate. In contrast to the
latter, the leading contribution to the dipole-dipole interaction potential is
to a high accuracy governed by the terms of order $\mathcal{O}(h_0^2)$, as
caused purely by the temporal modulation.

Evaluating the leading part of the dipole-dipole
interaction potential for a pair of particles with the
coordinates ${\mathbf r}_l=(x_l,y_l)$ and $l, l' \in \{1,2\}$, $l' \ne l$ (the elevation $z$ is fixed),
yields:
\begin{equation}
U_{\rm dd}({\mathbf r}_{12})=\alpha h_0^2 \left[ \frac{s_1}{r^3}
-\frac{s_2({\mathbf r}_{12}\cdot \hat{\mathbf e}_x)^2}{r^5} \right]\,
\label{Udd}
\end{equation}
with the time-dependent functions $s_1(\beta,t)=1+\beta\cos(4\pi \tilde f t)$ and $s_2(\beta,t)=(3/2)(1+\beta)(1+\cos(4\pi \tilde f t))$.
Here, $\hat{\mathbf e}_x=(1,0,0)$ is the unit vector
along the $x$ axis, ${\mathbf r}_{12}={\mathbf r}_1-{\mathbf r}_2$,
and $r=|{\mathbf r}_{12}|$ is the distance between the particles.

The deterministic dynamics of the pair of particles, including the
motion in the FGF potential, Eq.~\ref{eq-Us}, and the dipole-dipole
interactions as in Eq.~\ref{Udd}, obeys the equations:
%
\begin{equation}
\dot {\mathbf r}_l = F_{\rm s}\hat{\mathbf e}_x+\frac{\alpha h_0^2}{r^5}\left[\left(3 s_1 -5s_2\frac{x_{ll'}^2}{r^2} \right){\mathbf r}_{ll'}+ 2 s_2 x_{ll'}\hat{\mathbf e}_x \right], \label{eq-x1y1-det}
\end{equation}
where $F_{\rm s}(x_l,t)$ is the force exerted on dipole $l$ by the field of substrate,
see Eq.~\ref{Fs(x,t)}. In the case of no dipole-dipole interaction, $\alpha=0$, the dynamics
of particles reduces to the independent but identical one-dimensional
translation across the stripes, as described by Eqs.~\ref{dotx-eq-single-det}
and \ref{Fs(x,t)}, which admit no relative motion. The relative motion comes
into play when the particles start to interact, $\alpha>0$.

To describe the slow dynamics of interacting particles, we
perform the time-averaging of Eqn.~\ref{eq-x1y1-det}.
We note that in addition to fast evolving functions in $F_{\rm s}$ oscillating
with frequency $\tilde f$, the dipole-dipole interactions also excite
oscillations with the double frequency, $2\tilde f$, entering via the functions $s_1$ and $s_2$. This time dependence suggests the corresponding ansatz:
\begin{eqnarray}
{\mathbf r}_l(t)  ={\mathbf R}_l(t)+\mathbf{\delta r}_l(t),  \quad \mathbf{\delta r}_l =\mathbf{\delta r}_l^{(1)}+\mathbf{\delta r}_l^{(2)}\,,\\
\mathbf{\delta r}_l^{(1)}  =\tilde{\mathbf{r}}_l^{(1)}{\rm e}^{2\pi i \tilde f t}+\textrm{c.c.}, \quad \mathbf{\delta r}_l^{(2)}=\tilde{\mathbf{r}}_l^{(2)}{\rm e}^{4\pi i \tilde f t}+\textrm{c.c.}\,, \label{delta-r(2)}
\end{eqnarray}
where ${\mathbf R}_l=(X_l,Y_l)=\tilde f \int_0^{1/\tilde f} {\mathbf r}_l(t)\, dt$
denotes the solution averaged over the fast oscillatory timescales,
the superscripts ``$(1)$'' and ``$(2)$'' are used to mark the solutions
oscillating with the single ($\tilde f$) and double ($2\tilde f$)
frequency, respectively. The $\tilde{\mathbf r}_l^{(j)}$ stand for
the complex amplitudes and \textrm{c.c.} means the complex conjugate.
Note that the leading part of solution for
${\mathbf r}_l^{(1)}=(\tilde x(X_l),0)$ is determined by the earlier considered
case $\alpha=0$ with $\tilde x(X_l)$ given by $\tilde x(X_l)$
in Eq.~\ref{x-puls}.

Before we proceed to the derivation of the
complex amplitudes ${\mathbf r}_l^{(2)}$, we
expand all spatially dependent functions in Eqn.~\ref{eq-x1y1-det}
as $g({\mathbf r}_{12}) \approx g({\mathbf R}_{12})+\partial_{{\mathbf r}_{12}} g({\mathbf R}_{12}) \cdot \mathbf{\delta r}_{12}$.
Retaining the leading contributions, for the
evolution of the solution evolving with the double frequency we obtain the equations:
$\partial_t \boldsymbol{\delta}{\mathbf r}_l^{(2)}=\alpha h_0^2/R^5[(3 \tilde s_1 -5 \tilde s_2 X_{ll'}^2/R^2)\mathbf{R}_{ll'}+ 2\tilde s_2 X_{ll'}\hat{\mathbf e}_x]$. Here, $\tilde{s}_1=\beta\cos(4\pi \tilde f t)$ and
$\tilde{s}_2=(3/2)(1+\beta)\cos(4\pi \tilde f t)$ are the
quickly evolving parts of functions $s_1$ and $s_2$ oscillating with the double frequency,
$2\tilde f$. Using the exponential representation of the function $\cos(4\pi\tilde f t)$ and
taking into account the explicit temporal dependence in $\mathbf{\delta r}^{(2)}$,
see Eq.~\ref{delta-r(2)}, we solve the above equations
for the complex amplitudes to arrive at:
\begin{equation}
\tilde{\mathbf{r}}_l^{(2)} = -\frac{3 i \alpha h_0^2}{16 \pi \tilde f R^5} \left[ {\mathbf p}(\beta,{\mathbf R}_{ll'}) -5(1+\beta)\frac{X_{ll'}^2}{R^2}{\mathbf R}_{ll'}\right],\label{x2y2-puls}
\end{equation}
with ${\mathbf p}=(2(1+2\beta)X_{ll'},2\beta Y_{ll'})$.
From Eq.~\ref{x2y2-puls} for $\tilde y_l^{(2)}$ we see that oscillations along the stripes
of the FGF occur only if the particles have different $y$ coordinates,
$Y_{12}\ne 0$. For a pair of particles moving across the stripes one behind
another no oscillations transverse to the propagation direction takes place.

The relative contribution of the quickly oscillating solutions scales as:
$|\tilde{\mathbf r}_l^{(2)}|/|\tilde{\mathbf r}_l^{(1)}| \simeq \alpha h_0 {\rm e}^{2\pi z}/R^4$.
For our system, the enumerator can be of order $1$.
This means that when particles are widely separated, $R\gg 1$, the fast dynamics
corresponds to oscillations (around the time-averaged solution) with the
frequency $\tilde f$. As long as particles come closer, the relative amplitude of
oscillations with the double frequency increases and at separations about few diameters,
the fast dynamics presents the superposition of oscillations with both frequencies,
$\tilde f$ and $2 \tilde f$, around the slowly evolving state.

We are now ready to figure out the leading contributions into
the time-averaged equations. Taking into account the solutions
that determine the fast dynamics, we average over time Eqs.~\ref{eq-x1y1-det}
and arrive at the equations:
\begin{equation}
\dot {\mathbf R}_l = v_0\hat{\mathbf e}_x + \frac{\alpha h_0^2}{R^5}\left[\left(3 S_1 -5S_2\frac{X_{ll'}^2}{R^2} \right){\mathbf R}_{ll'}+2 S_2 X_{ll'}\hat{\mathbf e}_x \right], \label{eqs-XlY1}
\end{equation}
where $v_0$ is given by Eq.~\ref{v0}
and $S_1=\overline{s}_1=1$, $S_2=\overline{s}_2=(3/2)(1+\beta)$ are the time averaged
counterparts of the functions $s_1$ and $s_2$.

The time-averaged effect of dipole-dipole interaction of a pair of
particles is described by the effective potential:
\begin{equation}
U_{\rm dd}({\mathbf R}_{12})=\frac{\alpha h_0^2}{R^3} \left[ 1-\frac{3(1+\beta)}{2}\frac{X_{12}^2}{R^2}\right], \label{Udd-hf}
\end{equation}
where ${\mathbf R}_{12}={\mathbf R}_1-{\mathbf R}_2=(X_1-X_2, Y_1-Y_2)$
and $R=|{\mathbf R}_{12}|$. Alternatively, if we introduce the polar
angle $\vartheta$ such that ${\mathbf R}_{12}=R (\cos\vartheta, \sin\vartheta)$, then:
\begin{equation}
U_{\rm dd}(R,\vartheta)=\frac{\alpha h_0^2}{R^3} \left[ 1-\frac{3}{2}(1+\beta)\cos^2\vartheta\right]. \label{Udd-hf-theta}
\end{equation}

Finally, we note that the same effective potential,
Eq.~\ref{Udd-hf}, would follow from Eq.~\ref{Udd},
if we naively replaced functions $s_1$, $s_2$ and all
the coordinates by their time-averaged counterparts.
This result, however, is not obvious a priori, before
the order of magnitude of the oscillating contributions
is evaluated. We have also made a more careful analysis
of other time averaged contributions such as e.g., the effects
of the double frequency harmonics on the single particle motion
and of the substrate field on the dipole-dipole interaction potential.
The analysis shows that all these contributions
present only small corrections to
the leading one, as obtained in this section.

\section*{Acknowledgments}
We thank Tom H. Johansen for providing the FGF. A.S. and P.T. were supported via a bilateral German-Spanish program
funded by DAAD (project No. 57049473). P.T. further acknowledges support
 from the ERC starting grant ``DynaMO''
(No. 335040) and from the programs RYC-2011-07605, and FIS2011-15948-E.


\begin{thebibliography}{0}
\bibitem{note1}
The exact value of $\chi$ is difficult to estimate
since, in general, it depends on the magnetic doping of the paramagnetic colloids, which
can vary from one stock solution to another.
\end{thebibliography}
\end{document}